\documentclass{moriond}

\usepackage{graphicx}
\graphicspath{{./figs/}}

\usepackage{tabularx}
\usepackage{booktabs}
\usepackage{footnote}
\usepackage{threeparttable}  
\usepackage{multirow}
\usepackage{bm}
\usepackage{relsize}
\usepackage{gensymb}

\usepackage{chngcntr}

\usepackage{sidecap}

\usepackage{amsmath}
\usepackage{amssymb}

\def\be{\begin{equation}}
\def\ee{\end{equation}}
\def\bea{\begin{eqnarray}}
\def\eea{\end{eqnarray}}

\newcommand{\Photo}{}

\begin{document}
\vspace*{4cm}
\title{Nuclear modification factor of charged particles and light-flavour hadrons in p--Pb collisions measured by ALICE}

\author{GYULA BENCEDI (for the ALICE Collaboration)}

\address{Wigner Research Centre for Physics of the Hungarian Academy of Sciences\\
Instituto de Ciencias Nucleares, UNAM, Mexico City
}

\maketitle\abstracts{
The hot and dense strongly interacting Quark-Gluon Plasma (sQGP) created in ultra-relativistic heavy-ion collisions can be probed by studying high-$p_{\rm T}$ particle production and parton energy loss. Similar measurements performed in p-Pb collisions may help in determining whether initial or final state nuclear effects play a role in the observed suppression of hadron production at high-$p_{\rm T}$ in Pb--Pb collisions. By examining the nuclear modification factors through the comparison of identified hadron yields in different collision systems one can gain insight into particle production mechanisms and nuclear effects.}


\section{Introduction}\label{sec:1}

A Large Ion Collider Experiment~\cite{Abelev:2014ffa} (ALICE) at the LHC is dedicated to study the strongly interacting deconfined medium of quarks and gluons, the Quark-Gluon Plasma (QGP), created in ultra-relativistic heavy-ion collisions~\cite{Bass:1998vz}. At the early stage of the collision some partons experience scatterings with large momentum transfer while propagating through the created medium and they lose energy. In Pb--Pb collisions this leads to a suppression
of high-$p_{\rm T}$ particles in the final state (via jet quenching~\cite{Gyulassy:1999zd,Levai:2001dc}) with respect to the hadron yields expected in a scenario of incoherent superposition of pp collisions. The measured suppression of charged hadrons is larger than observed at RHIC~\cite{PhysRevLett.88.022301} due to the higher energy density reached at the LHC. In this context, p--Pb collisions can be used as a control experiment to establish whether initial state effects play a role in the observed suppression of hadron production in Pb--Pb collisions.\\
\indent In order to study the suppression and to disentangle hot (QGP specific) and cold (e.g. Cronin-enhancement~\cite{Antreasyan:1978cw}) nuclear matter effects, the nuclear modification factors ($R_{\rm pPb}$, $R_{\rm AA}$) are introduced: 
\begin{equation*}
R_{\rm pPb} = \frac{\text{d}^{2}N_{\text{pPb}}/\text{d}y\text{d}p_{\rm T}}{\left<T_{\rm pPb}\right>\text{d}^{2}\sigma^{\rm INEL}_{\rm pp}/\text{d}y\text{d}p_{\rm T}}~,~R_{\rm AA} = \frac{\text{d}^{2}N_{\text{AA}}/\text{d}y\text{d}p_{\rm T}}{\left<T_{\rm AA}\right>\text{d}^{2}\sigma^{\rm INEL}_{\rm pp}/\text{d}y\text{d}p_{\rm T}}~,
\end{equation*}
where $N_{\text{pPb}}(N_{\text{AA}})$ and $\sigma^{\rm INEL}_{\rm pp}$ represent the particle yield in p--Pb (Pb--Pb) and the inelastic cross-section in pp collisions, respectively. In p--Pb (similar to Pb--Pb), the nuclear overlap function is defined as $\left<T_{\rm pPb}\right> = \left<N_{\rm coll}\right> / \sigma^{\rm NN}_{\rm INEL}$. It is determined from the Glauber model~\cite{Miller:2007ri} and proportional to the average number of binary nucleon-nucleon collisions $\left<N_{\rm coll}\right>$. In the absence of nuclear effects the $R_{\rm pPb}$ is expected to be equal to unity.\\
\indent ALICE has excellent particle identification capabilities (PID) in the central barrel ($|\eta| < 0.9$). The two main tracking detectors, the Inner Tracking System (ITS) and the Time Projection Chamber (TPC) allow for PID via measurement of the specific energy loss. The Time Of Flight (TOF) detector and the High Momentum Paricle Identification Detector (HMPID) identify particles by measuring the time-of-flight and the Cherenkov angle, respectively.\\
\indent These proceedings give an overview of recent ALICE results on the nuclear modification factors for charged as well as identified light-flavour hadrons in Pb--Pb and p--Pb collisions.


\section{$R_{\rm AA}$ and $R_{\rm pPb}$ for charged particles}\label{sec:2}

\begin{figure}[!htbp]
  \hspace{0.05\linewidth}
\begin{minipage}[c]{0.4\linewidth}
  \centerline{\includegraphics[width=1.05\linewidth]{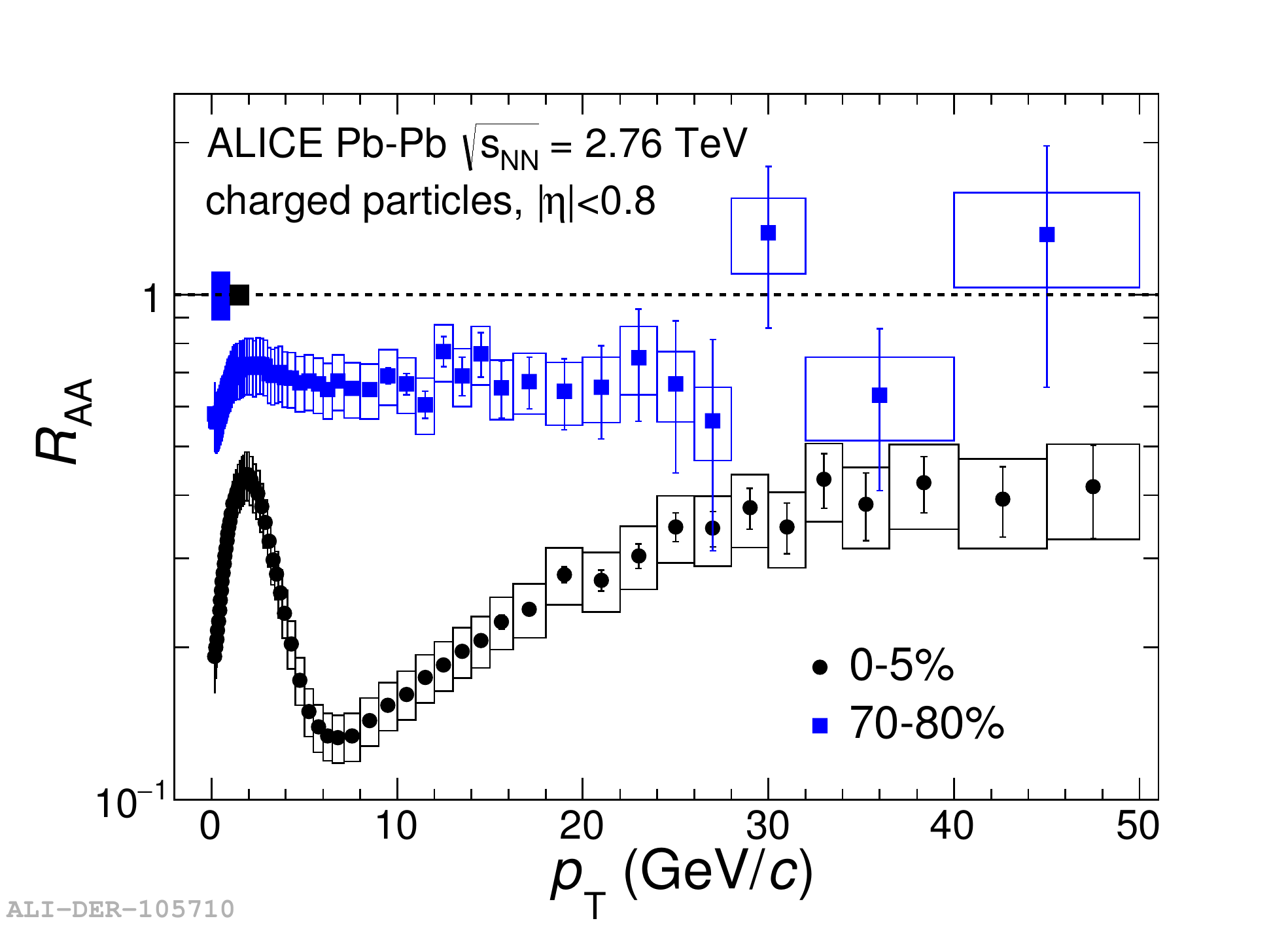}}  
  \centerline{\includegraphics[width=\linewidth]{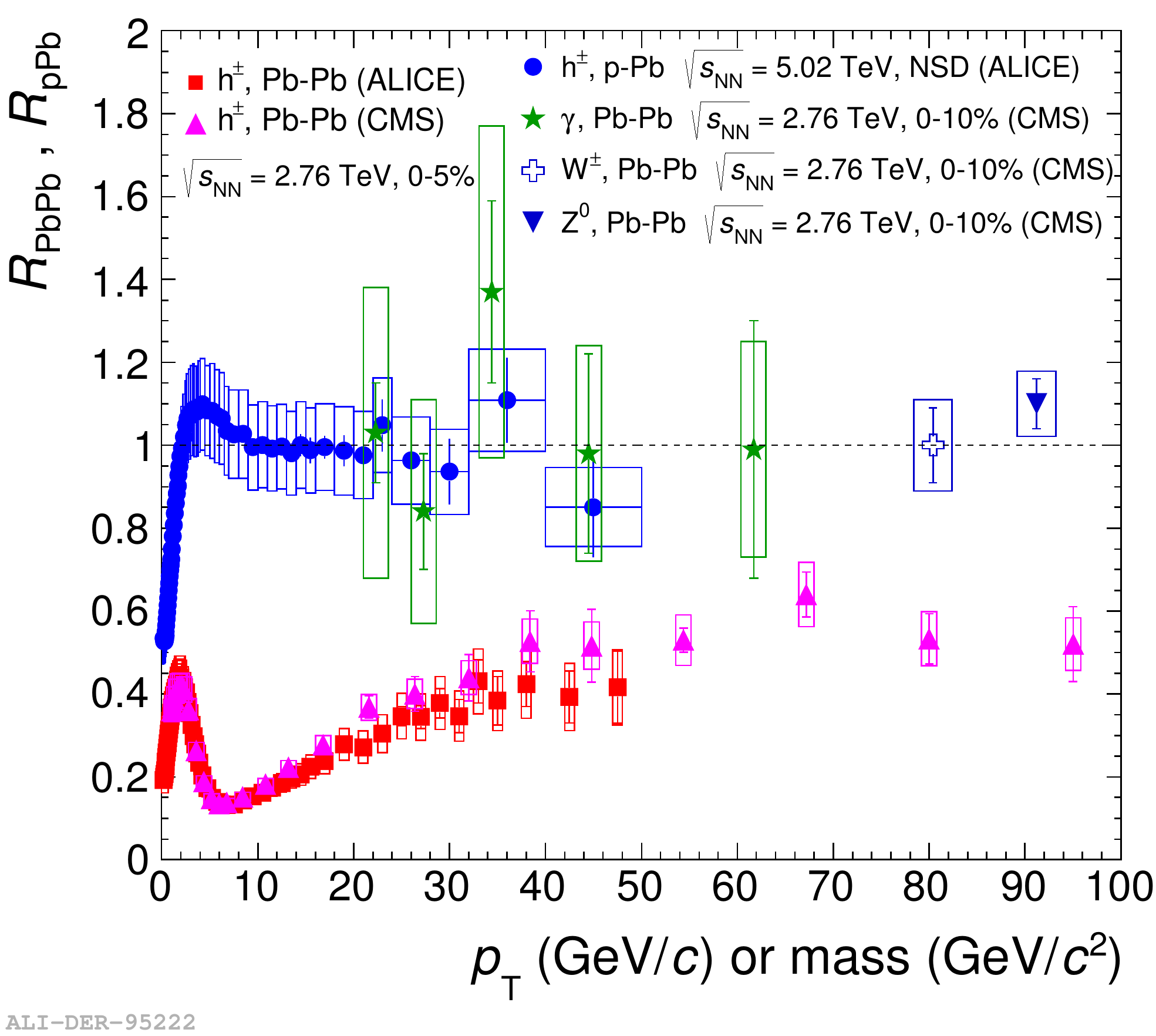}}
\end{minipage}
  \hspace{0.05\linewidth}
\begin{minipage}[c]{0.44\linewidth}
\centerline{\includegraphics[width=\linewidth]{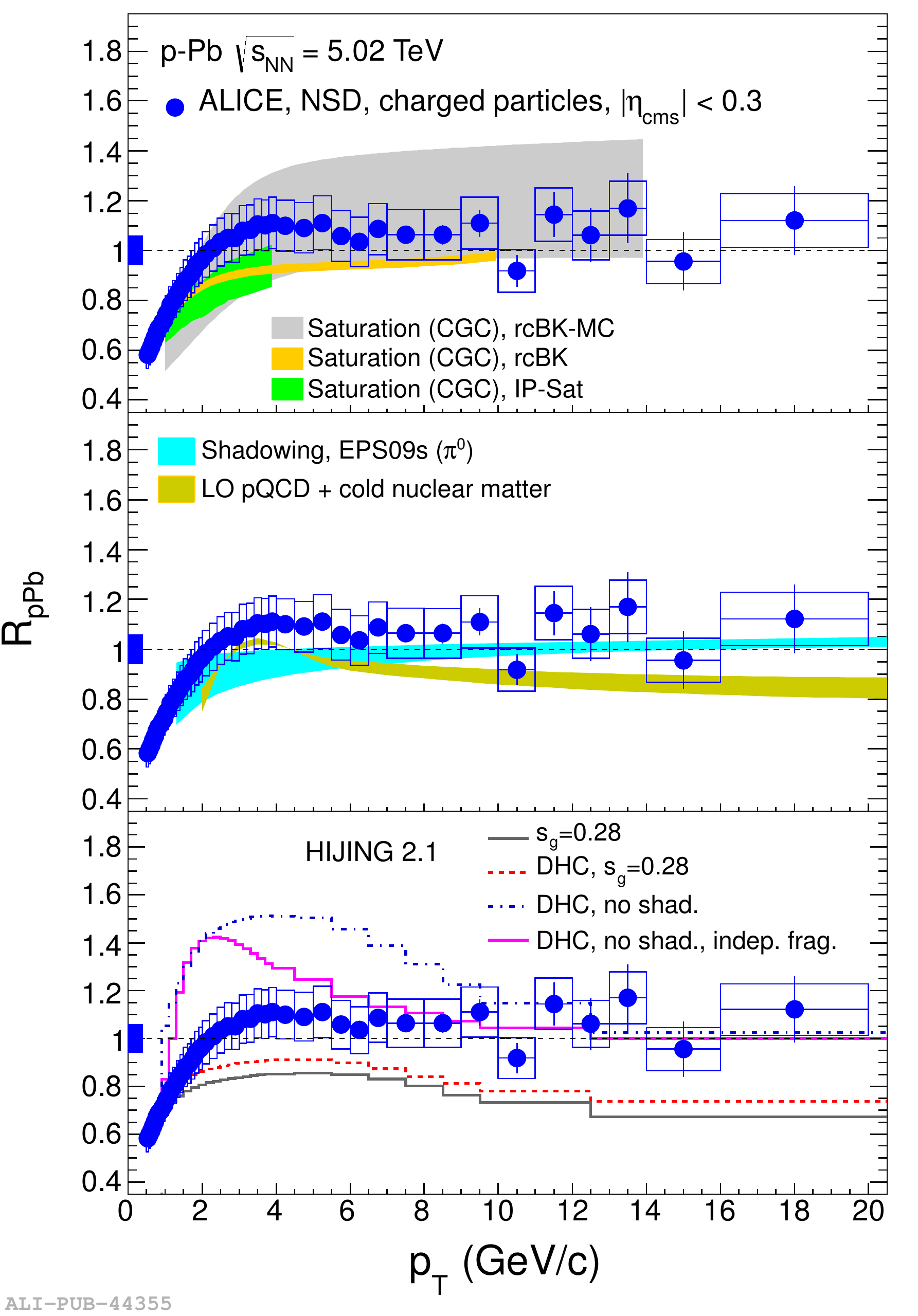}}
\end{minipage}
  \caption[]{$R_{\rm AA}$ and $R_{\rm pPb}$ of charged particles. (Top left) $R_{\rm AA}$ is shown in central (0-5\%) and peripheral (70-80\%) Pb--Pb collisions. (Bottom left) Comparisons of $R_{\rm AA}$ and $R_{\rm pPb}$ measured by ALICE and CMS. (Right) $R_{\rm pPb}$ from ALICE for $|\eta_{\rm cms}| < 0.3$ (symbols) are compared to model calculations (bands or lines).
  }
\label{fig:sec12}
\end{figure}

The nuclear modification factor, $R_{\rm AA}$, in Pb--Pb collisions at $\sqrt{s_{\rm NN}} = 2.76 \text{\,TeV}$ is shown for two centrality intervals~\cite{Abelev:2012hxa} in the top left panel of Fig.~\ref{fig:sec12}. There is a significant suppression of charged hadron yields for the most central (0-5\%) collisions. The nuclear modification factor exhibits a minimum at around $p_{\rm T} = $ 6-7~GeV/$c$ and a significant rise for $p_{\rm T} > 7$~GeV/$c$, indicating a reduction of the relative energy loss. For peripheral collisions (70-80\%) only a moderate suppression and a weak $p_{\rm T}$ dependence is observed. 

The observed suppression at a given centrality is a consequence of the interplay of the parton $p_{\rm T}$ distribution, the medium density and the gluon-to-quark ratio. The relative contributions of these effects can be studied in comparison to models. In general, all the models capture the rise of $R_{\rm AA}$ due to a decrease of the relative energy loss with increasing $p_{\rm T}$, but there are remarkable quantitative deviations in some cases.\\
\indent In the bottom left panel of Fig.~\ref{fig:sec12} the nuclear modification factor, $R_{\rm pPb}$, in p--Pb collisions at $\sqrt{s_{\rm NN}} = 5.02 \text{\,TeV}$ is shown for charged particles~\cite{ALICE:2012mj}, in comparison with $R_{\rm AA}$ for most central (0-5\%) collision measured by ALICE and CMS. Moreover, comparisons are also shown for particles which are not sensitive to QCD dynamics (direct photon, W$^{\pm}$, Z$^{0}$) measured by CMS.\\
For $p_{\rm T} \gtrsim 2$ the $R_{\rm pPb}$ is consistent with unity showing that the large suppression observed for $R_{\rm AA}$ at high-$p_{\rm T}$ is related to the jet quenching in QGP and not to inital state effects. 

\indent A comparison of the p--Pb data to models is crucial for the understanding of cold nuclear matter effects. In the right panel of Fig.~\ref{fig:sec12} $R_{\rm pPb}$ for $|\eta_{\rm cms}| < 0.3$ is compared to theoretical predictions. Some predictions based on the Colour Glass Condensate (CGC) model are consistent with the measurement within uncertainties. Leading order (LO) pQCD calculations incorporating cold nuclear matter effects underpredict the data at high $p_{\rm T}$, while the shadowing calculations based on NLO with EPS09s PDF's and DSS fragmentation functions describe the data well for $p_{\rm T} > 6$~GeV/$c$. The HIJING 2.1 model (with shadowing) describes the trend observed in the data.


\section{$R_{\rm AA}$ and $R_{\rm pPb}$ for identified light-flavour hadrons}\label{sec:3}

The nuclear modification for identified particles gives more details about the in-medium interactions of partons fragmenting into hadrons due to the different color Casimir factors of quarks and gluons.

\begin{figure}[!htbp]
\begin{minipage}[c]{\textwidth}
  \centerline{\includegraphics[width=0.6\linewidth]{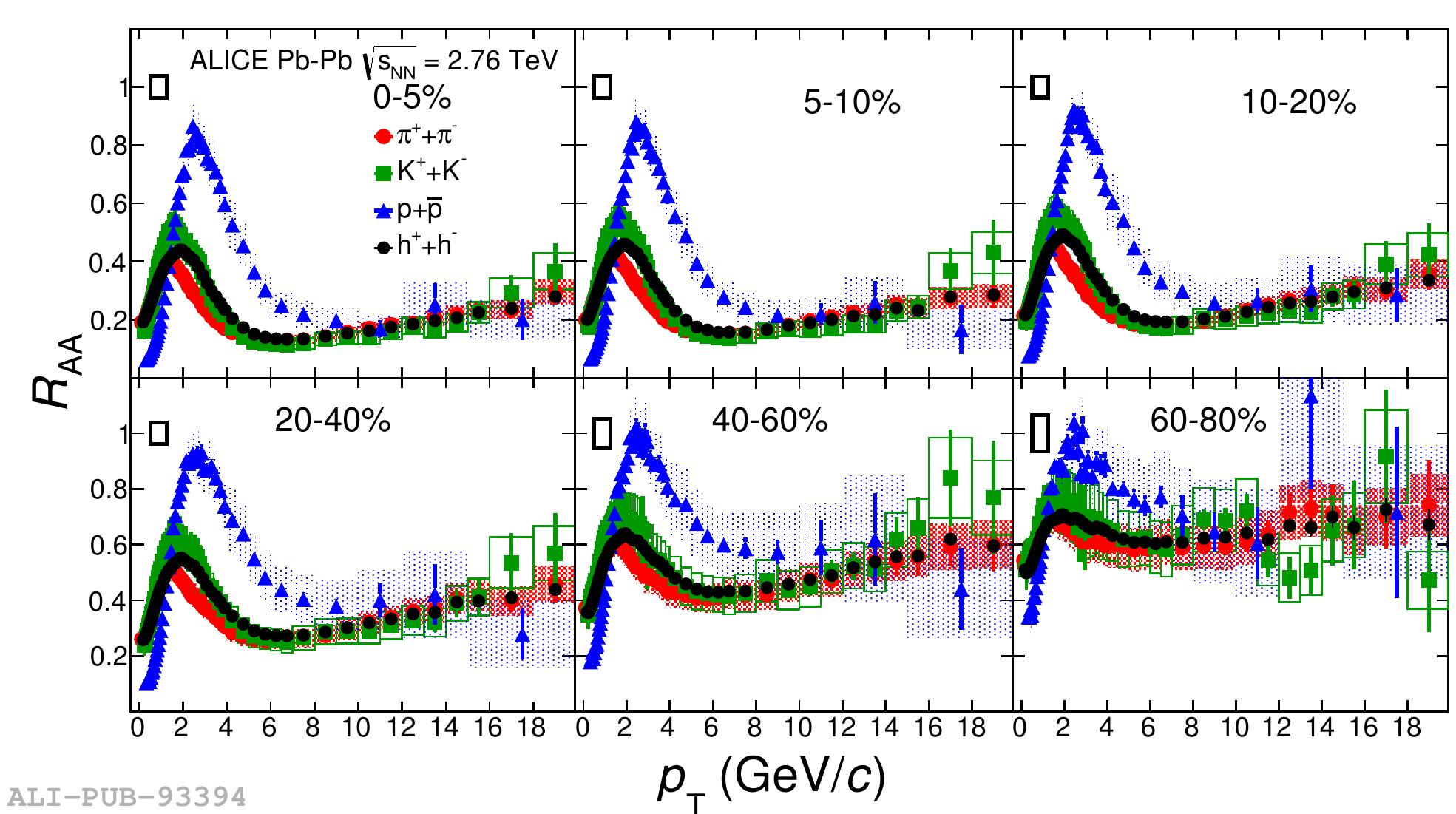}}
\end{minipage}
  \caption[]{$R_{\rm AA}$ for different particle species in six centrality intervals in Pb--Pb collisions.}
\label{fig:sec21}
\end{figure}

\indent Fig.~\ref{fig:sec21} shows the $R_{\rm AA}$ for $\pi^{\pm}$, K$^{\pm}$ and p($\bar{\rm p}$) for six centrality classes in Pb--Pb collisions~\cite{Adam:2015kca}. Results show that all particle species are equally suppressed for $p_{\rm T} > 10$ GeV/$c$. This indicates that particle ratios are similar to those of jets in the vacuum. Also, jet quenching does not have effect on the particle species composition of the hard core of the quenched jet.\\
At lower $p_{\rm T}$ ($< 10$ GeV/$c$) protons are less suppressed for all centralities. This mass dependence of the nuclear modification of the $p_{\rm T}$ spectra~\cite{Abelev:2014laa} is consistent with radial flow.

\begin{figure}[!htbp]
\begin{minipage}{0.4\linewidth}
  \centerline{\includegraphics[width=\linewidth]{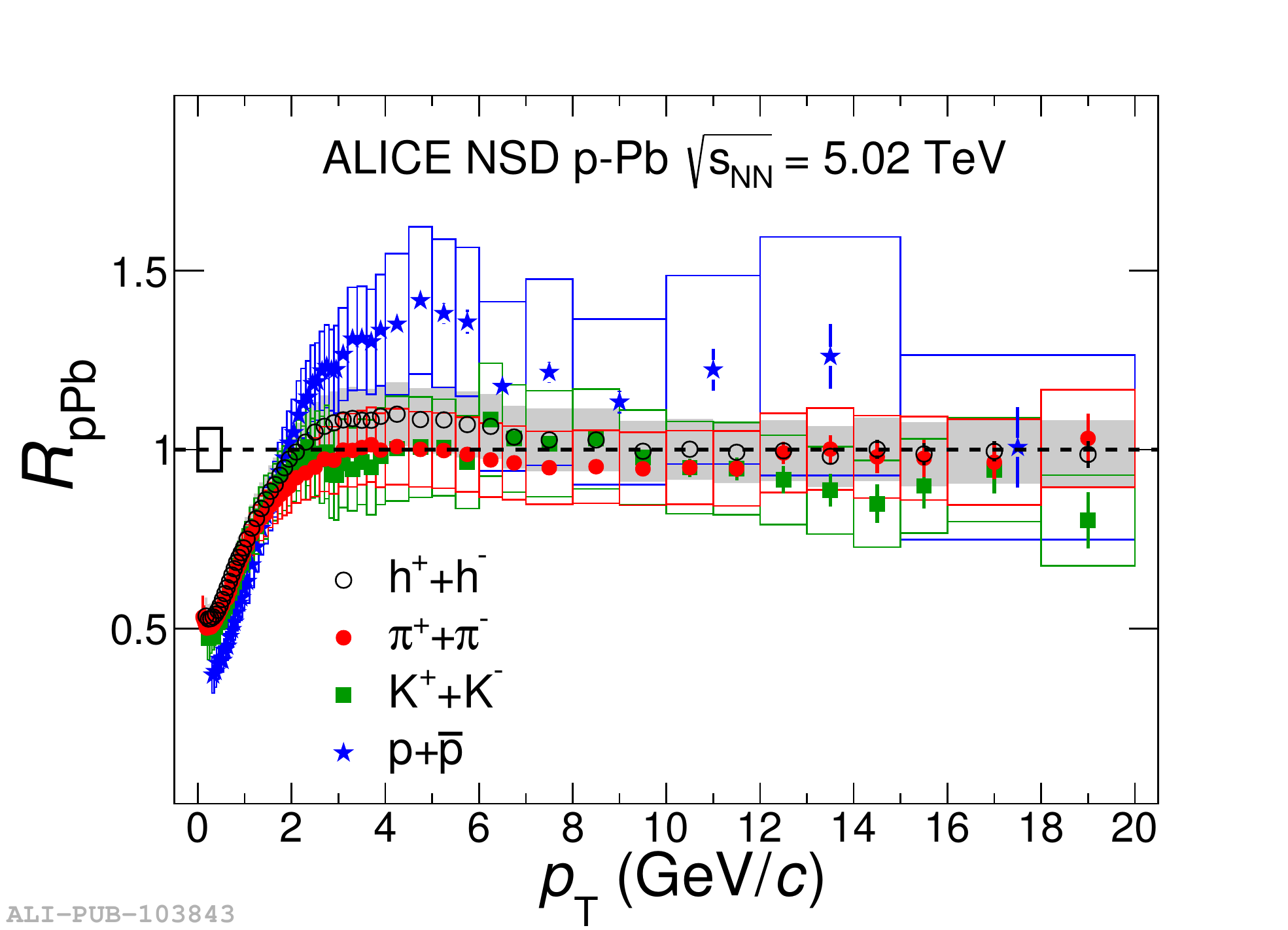}}
\end{minipage}
\hfill
\begin{minipage}{0.6\linewidth}
  \centerline{\includegraphics[keepaspectratio, width=\columnwidth]{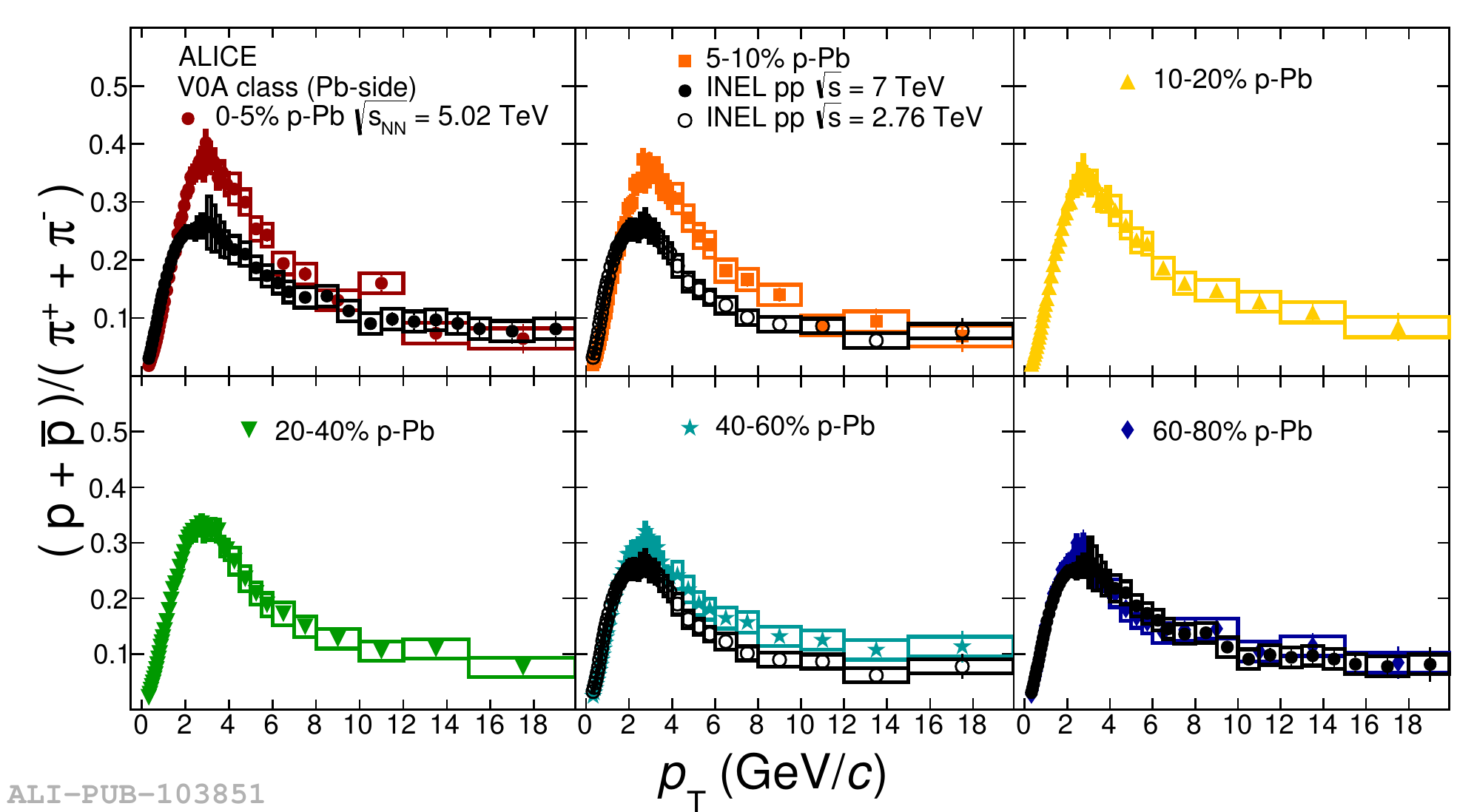}}
\end{minipage}
  \caption[]{(Left) $R_{\rm pPb}$ for different particle species in NSD p--Pb collisions. (Right) Proton-to-pion ratios for different V0A multiplicity classes in p--Pb collisions compared to pp data.
  }
\label{fig1:sec22}
\end{figure}

\indent The left panel of Fig.~\ref{fig1:sec22} shows the identified hadron $R_{\rm pPb}$ compared to that for inclusive charged particles in non-single diffractive (NSD) p--Pb events~\cite{Adam:2016dau}. The nuclear modification factors at high $p_{\rm T}$ ($> 10$ GeV/$c$) are consistent with unity within uncertainties, suggesting that final state effects do not play a role. Around $p_{\rm T} = 4$ GeV/$c$ the $R_{\rm pPb}$ for protons is above unity. This enhancement is $\sim$3 times larger than that for charged hadrons. Conversely, for charged pions and kaons the enhancement is below that of charged particles. A similar enhancement of protons was seen in Pb--Pb data and a similar pattern was observed at RHIC.\\
Since the calculation of $\left<T_{\rm pPb}\right>$ is technically more intricate for the smaller p--Pb system~\cite{Adam:2014qja}, spectra modification is studied via the multiplicity dependence of the proton-to-pion ratio in the right panel of Fig.~\ref{fig1:sec22}. For $p_{\rm T} < 10$ GeV/$c$ a strong multiplicity dependence is visible, which is qualitatively similar to that observed in Pb--Pb collisions by ALICE. Conversely, at high $p_{\rm T}$ the trends are similar to pp measurements at $\sqrt{s} = 2.76 \text{\,TeV}$ and at $\sqrt{s} = 7 \text{\,TeV}$.


\section{Summary}

\indent The $R_{\rm AA}$ of charged hadrons shows suppression for all centralities, with the strongest effect for the most central (0-5\%) case and decreasing towards higher $p_{\rm T}$ with the decrease of the relative parton energy loss. On the other hand, the $R_{\rm pPb}$ follows binary collision scaling at high $p_{\rm T}$ indicating that the strong suppression in Pb--Pb is due to final state effects.\\
\indent For light-flavoured particles, the $R_{\rm AA}$ shows no particle species dependence at high $p_{\rm T}$ within uncertainties, which points out the fact that jet quenching does not produce large particle species dependent effects in the hard core of the jet, and that all fragments lose energy coherently. At lower $p_{\rm T}$, protons are less suppressed for all centralities and the mass dependence is a reflection of radial flow. At the same time, the $R_{\rm pPb}$ for protons shows a moderate Cronin enhancement at intermediate $p_{\rm T}$, while pions and kaons are rather lower than charged hadrons, which indicates little or no nuclear modification. At high $p_{\rm T}$, the yields of species are consistent with binary collision scaling.

\section*{Acknowledgments}

This work was supported by CONACYT under the grant No. 260440, by DGAPA-UNAM under PAPIIT grant IA102515 and by the Hungarian Research Fund (OTKA) under contract NK 106119, K 104260, TET 12 CN-1-2012-0016. The EPLANET program supported the mobility between Latin America and Europe.

\end{document}